\documentclass{sigchi}

\usepackage{balance}  
\usepackage{graphics} 
\usepackage{times}    
\usepackage[hyphens]{url}      
\usepackage[usenames,dvipsnames]{xcolor} 
\usepackage{tikz}
\usetikzlibrary{arrows}
\usepackage{booktabs}
\usepackage{pgfplots}
\usepackage{tabularx}

\makeatletter
\def\url@leostyle{%
  \@ifundefined{selectfont}{\def\UrlFont{\sf}}{\def\UrlFont{\small\bf\ttfamily}}}
\makeatother
\def\pprw{8.5in}
\def\pprh{11in}

\usepackage[pdftex]{hyperref}
\hypersetup{
pdftitle={Structure and Dynamics of Coauthorship, Citation, and Impact within CSCW},
pdfauthor={LaTeX},
pdfkeywords={SIGCHI, proceedings, archival format},
bookmarksnumbered,
pdfstartview={FitH},
colorlinks,
citecolor=black,
filecolor=black,
linkcolor=black,
urlcolor=black,
breaklinks=true,
}

\newcommand{\specialcell}[2][l]{%
  \begin{tabular}[#1]{@{}l@{}}#2\end{tabular}}

\begin{document}

\title{Structure and Dynamics of Coauthorship, Citation, and Impact within CSCW}

\numberofauthors{4}
\author{
  \alignauthor Brian C. Keegan\\
    \affaddr{Northeastern University}\\
    \affaddr{360 Huntington Ave.\\Boston, MA, USA}\\
    \email{b.keegan@neu.edu}\\
  \alignauthor Dan Horn\\
    \affaddr{Booz Allen Hamilton}\\
    \affaddr{8283 Greensboro Dr.\\McLean, VA 22102}\\
    \email{danhorn@gmail.com} 
  \alignauthor Thomas A. Finholt\\
    \affaddr{University of Michigan}\\
    \affaddr{1075 Beal Ave.\\Ann Arbor, MI 48109}
    \email{finholt@umich.edu}
  \alignauthor Joseph `Jofish' Kaye\\
    \affaddr{Yahoo! Labs}\\
    \affaddr{701 1st Ave\\Sunnyvale, CA}
    \email{jofish@jofish.com}
}

\maketitle

\begin{abstract}
CSCW has stabilized as an interdisciplinary venue for computer, information, cognitive, and social scientists but has also undergone significant changes in its format in recent years. This paper uses methods from social network analysis and bibliometrics to re-examine the structures of CSCW a decade after its last systematic analysis. Using data from the ACM Digital Library, we analyze changes in structures of coauthorship and citation between 1986 and 2013. Statistical models reveal significant but distinct patterns between papers and authors in how brokerage and closure in these networks affects impact as measured by citations and downloads. Specifically, impact is significantly influenced by structural position, such that ideas introduced by those in the core of the CSCW community (e.g., elite researchers) are advantaged over those introduced by peripheral participants (e.g., newcomers).  This finding is examined in the context of recent changes to the CSCW conference that may have the effect of upseting the preference for contributions from the core.  
\end{abstract}




\section{Introduction}
The genesis of computer-supported cooperative work (CSCW) as a field was in response to the simultaneous challenges of the emergence of personal computers and of computer networks~\cite{grudin_computer-supported_1994,grudin_taxonomy_2012}.  Connected systems introduced the possibility of applications designed to meet the needs of groups and organizations, and hence a novel context for examination of computer use and users.  Understanding this new context required insight at the intersection of computer science, information science, cognitive science, and social science.  At the beginning, then, CSCW was a highly interdisciplinary field where initial participants came from other domains and conferences were characterized by the ``sounds of ideologies clashing.''
  
CSCW has changed significantly since the early days. Thematically, the community has shifted from a focus on workplace group collaboration (``groupware'') towards analysis of complex and large-scale collaborative socio-technical systems (``social computing''). Methodologically, a focus on system building and evaluation has been replaced by quantitative and qualitative analyses of existing systems.  The CSCW community has coalesced into an independent discipline where researchers can establish their identity largely within the boundaries of the community, rather than bringing an outside identity with them (as the founders of the field had to do). Finally, CSCW has also undergone major structural changes moving from a biennial to an annual format, adopting a hybrid ``revise and resubmit'' review format, and eliminating page limits that aresignificant departures from prevailing publication models in computer science and may systematically alter the composition of the community and its scholarship.

An important question to ask about the transition of CSCW from a nascent to an established field is the effect of this transition on the creation and impact of ideas.  In particular, has the community retained openness to newcomers and new ideas?  Or, as is the case in many maturing fields, has structural position become disproportionately important in the visibility and uptake of ideas?  Historically, addressing questions of this type has been a monumental enterprise.  Today, however, extensive bibliographic databases with accompanying metadata have introduced a revolution in the ability to characterize and analyze the structure of research communities.  While some fields have embraced this revolution with enthusiasm (e.g., physics), CSCW researchers have been relatively quiet, with only a few attempts to understand the structure of the CSCW community.  A primary contribution of this paper, then, is a comprehensive depiction of the co-authorship and citation networks formed among CSCW researchers.  Through examination of these networks we will
\begin{enumerate}
\item Produce a representation of the structure of the CSCW community, including visualizations of this structure;
\item Explore hypotheses about the role of structural position on the impact of ideas (e.g., as measured through citation and download frequency); and
\item Describe the evolution of the CSCW community over time in terms of cohesion and broader relevance.
\end{enumerate}

\section{Background and Approach}
\subsection{Bibliometrics and complex networks}
Scientometric and bibliometric approaches to understanding the evolution of scientific ideas and communities have historically relied heavily on the collection and analysis of citation patterns in scientific publications. Citation and coauthorship relationships capture distinct structures of scientific collaboration and knowledge production. Quantitative analyses of the patterns or regularities in these relationships have used methods from social network analysis to understand how social structures influence individuals' success, often more significantly than their personal attributes. Specifically, the position of actors or papers within a network enhances or constrains their access to resources that are explicitly needed for scientific production such as funding, data, and skills as well as implicit resources such as social capital and prestige.

Papers are the outputs from collaborative work integrating individuals' different skills, resources, and expertise, but they also require a shared understanding of the problem and domain. Coauthorship is a highly visible indicator of scientific collaboration as it encodes authors' interactions and collective accomplishment of their joint research. The accumulation of coauthorship relationships between authors reflects latent relationships such as trust, exchange, and shared norms that are crucial to the development of social capital.

Similarly, citations also signal the importance of a paper or author. Scientific research is cumulative and repeated references to prior papers and their authors provide another form of social capital indicating the value of these past contributions to on-going research. Although citations often occur in the absence of direct social relationships, the accumulation of citations can coalesce into larger structures that reveal the common theories and methods that form the foundation or define the identity of a field.

Scholarship using bibliometric data about coauthorship and citation goes back fifty years to early concepts of scale-free network degree distributions and models of preferential attachment~\cite{de_solla_price_networks_1965,de_solla_price_general_1976}. More recently, work by physicists and computer scientists re-invigorated the study of large-scale scientific citation and co-authorship patterns as exemplars of complex networks~\cite{martin_coauthorship_2013,moody_structure_2004,newman_structure_2001,newman_coauthorship_2004}. These networks are complex because constituent nodes have heterogeneous patterns of connectivity and clustering, their position in the network varies over time as other nodes enter and leave, and the intensity and direction of relationships can vary. Various methods and metrics have been applied to characterize influential scholars ~\cite{bornmann_what_2007,hirsch_index_2005}, overlaps in coauthorship and citation networks~\cite{ding_scientific_2011,martin_coauthorship_2013,wallace_small_2012}, the relationships between fields~\cite{bollen_clickstream_2009}, and changes in fields over time~\cite{herrera_mapping_2010,wuchty_increasing_2007}.

Bibliometric analyses of networks in the CSCW and CHI communities have built upon these approaches. A 2004 analysis of the CSCW community identified high levels of volatility in membership, strong ties to the larger HCI community, diversity in works cited across conferences and journals, and persistence of elite researchers' position over time~\cite{horn_six_2004}. Analysis of citation patterns within the CSCW community has suggested the presence of disparate cores reflecting distinct research topics and approaches involving technical system building versus qualitative user studies~\cite{jacovi_chasms_2006}. Geographic location is also associated with differences in the likelihood that a paper is to be tentatively accepted during the review process~\cite{fussell_international_2008}. Analyses of the larger CHI community have examined differences in representation and impact across countries and institutions~\cite{bartneck_scientometric_2009} as well as author persistence and gender diversity over time~\cite{cohoon_gender_2011,kaye_statistical_2009}. These prior analyses have provided a variety of valuable descriptive results, but they have not statistically tested whether network position creates social capital within the community.

\subsection{Network features of social capital}
The success of authors and their papers may be attributable to the resources (capital) authors have obtained such as having access to unique data (informational capital), receiving funding (financial capital), possessing specialized skills (human capital), as well as the relationships authors develop with each other (social capital). While definitions of social capital vary, it generally refers to the ability for individuals to secure other resources (information, money, skills, other social connections) by virtue of their membership and position within a social structure. The accumulation of social capital has been cited as a crucial factor in explaining success of individuals across diverse domains~\cite{burt_brokerage_2005,putnam_bowling_2001}. 

Social capital is not a new topic to CSCW researchers who have evaluated it within online communities like Facebook~\cite{burke_social_2011,jung_favors_2013}, but as a theoretical concept it has not been extended to collaborations of researchers within the CSCW community. We build upon this theoretical paradigm and use methods from network analysis (rather than traditional survey methods) to argue that the impact of research articulated through CSCW papers depends critically upon the social capital of its authors rather than the intrinsic value of its ideas. We describe four general network features that could generate social capital in the context of scientific coauthorship that will motivate descriptive and statistical analysis in subsequent sections: bridging, bonding, assortativity, and core membership.

\subsubsection{Bridging and bonding social capital}
The particular structural mechanisms that generate this social capital include possessing many contacts~\cite{bourdieu_forms_1986}, having strong ties~\cite{granovetter_strength_1973,krackhardt_strength_1992}, bridging structural holes~\cite{burt_network_2000,burt_brokerage_2005}, and being embedded within dense communities~\cite{coleman_social_1988}. Although each of these mechanisms has distinct network characteristics that can be empirically tested, the latter two mechanisms (also known as ``bridging'' and ``bonding'') have been the focus of the most on-going empirical work and theoretical debate~\cite{burt_brokerage_2005,putnam_bowling_2001}. In the context of scientific collaboration, both bridging and bonding mechanisms could contribute to the development of individuals' social capital. The value of this social capital can be captured in a systematic manner by examining the value the community assigns to an author through citations and downloads of their work. 

Node-level social network measures are indicators of authors' position and influence within collaborations as well as the role they play in the context of their scientific community. Bridging capital can create advantage for an author by making disconnected co-authors rely upon the brokering author to exchange knowledge. This gives the broker access to non-redundant and diverse pools of knowledge they can use to enhance their own research. Bonding capital can create advantage by embedding an author within a sub-group that has a history of being active within the community as well as frequently collaborating and sharing resources that the individual can then use to enhance their own research. We propose hypotheses detailing these processes in subsequent sections.

\subsubsection{Assortative and core membership social capital}
Other network features like assortativity and core community membership may also play a role in the development of social capital~\cite{ahuja_genesis_2011,rivera_dynamics_2010}. Intuitively, accomplished authors collaborating with other accomplished authors are more likely to produce high-impact work than novice authors collaborating with other novice authors. Moreover, these patterns of preferential degree mixing (assortativity) could also contribute to secondary effects such as stratification whereby the well-connected authors exclusively work with and cite other well-connected authors. Formally, the average connectivity of an author's neighbors reflects indirect access to various forms of capital that can enhance an individual's research.

The networks of coauthorship and citations may also be fragmented among various fault lines related to disciplinary approaches as well as institutional or geographic boundaries that prevent authors from working with or citing each other. This fragmentation manifests itself in the network as distinct components that are completely unconnected from each other. Some of these individual components may encompass a substantial fraction of total activity reflecting the core community of elite authors and papers but also excludes other authors and their work. Membership in this largest component may grant authors access to more of the implicit resources such as prestige that can enhance the impact of their work. However, this stratification of activity into unconnected components may also be deleterious to the long-term health of the scholarly community if it penalizes the impact of authors and papers who cannot access the core community..

\subsection{Approach}
Our goals are three-fold. First, we characterize the large-scale structural patterns of coauthorship and citation networks within CSCW to understand the connectivity of these networks. Visualizations of these large-scale networks as well as analyses of patterns in their connections demonstrate substantial similarities with other scientific fields and motivate subsequent research questions.

Second, we analyze changes in the activity and structural properties of the CSCW coauthorship and citation networks over time. The previous section identified network features such as brokerage, clustering, assortativity, and core membership as crucial mechanisms for generating social capital. The variance of these trends over time speaks both to the shifting contexts of collaboration among authors as well as the overall coherence of the field.  

Finally, we estimate statistical models for both authors and papers across these networks to understand how structural position influences their impact as measured by both citations and downloads. We test hypotheses related to the four network features of social capital discussed in the previous section. While we find evidence authors and papers both benefit from brokering and being connected to well-connected alters, authors benefit more from membership in the core community while papers benefit more from being embedded among other papers.

\section{Methods}
In addition to citation and co-authorship data used in previous studies, we also employ other metrics of impact such as downloads and citations having additional ecological validity.

\subsection{Data}
Through the use of custom web scraping scripts, we downloaded the publication meta-data for every work listed in the online tables of contents for the proceedings of every CSCW conference from 1986 through 2013.  Although works were primarily conference papers, these tables of contents also include short papers, posters, plenary sessions, panels, workshops, and tutorials. Because the ACM DL does not reliably distinguish these types of works, we are unable to exclude non-paper records from our analyses.

For each record, we extracted data that included the title and abstract of the work, the list of authors and their affiliations, the year of the conference, and the list of works cited by and citing the target work.  We refer to this base set of records as the CSCW Dataset For those works cited by the target paper, if the curators of the ACM DL were able to match the text of a citation with another record in the DL, we collected the ID of the cited work.  The ACM DL only includes a work that cites the target work if the citing work is also indexed in the DL.  This means that while only some of the works cited by the target document have ID numbers, all works citing the target document have ID numbers.

Because we wanted to be able to understand the citation networks involving CSCW publications, we captured a second dataset, which we refer to as the Extended Citation Dataset.  The Extended Citation Dataset has 12,365 records including all works indexed by the ACM DL that cite or are cited  by CSCW papers that were not already captured in the CSCW Dataset.  This includes journal articles, conference papers, books and other works that reference or are referenced by works presented at CSCW conferences.  The publication metadata for these works were collected in a fashion identical to what was done for the CSCW Dataset.

One challenge in analyzing bibliometric databases is that of author name disambiguation~\cite{smalheiser_author_2009}. Essentially, there are two main challenges in author name disambiguation ? distinguishing multiple authors who may publish using the same name, and combining multiple variants of a single author who is listed under different names. In our approach, we rely on the name disambiguation efforts that the maintainers of the ACM DL have implemented.  Each author of a paper in the ACM DL has an author ID taking into account some name variants, for example based on ACM DL disambiguation, one author had 6 variants: 'Jeffery T. Hancock', 'Jeffrey T. Hancock', 'Jeff T. Hancock', 'Jeffrey Hancock', 'Jeff Hancock', and 'Jeffery Hancock'.

There were a total of 1,288 publications associated with the 16 published proceedings between from 1986 through 2013. For each paper, there are author data such as names and institutional affiliation, scientometric data for authors such as downloads and publications, scientometric data for papers such as downloads and citations, and bibliometric data for works referenced by the paper as well as works citing the paper. Because the ACM DL generally only assigns publication IDs to ACM-affiliated proceedings, references and citations to papers outside the ACM do not have unique identifiers and are omitted from our analysis, leaving 11,080 publications with valid ACM DL IDs.

\subsection{Network Construction}

\begin{figure*}[!htb]
\minipage{0.45\textwidth}
  \includegraphics[width=\linewidth]{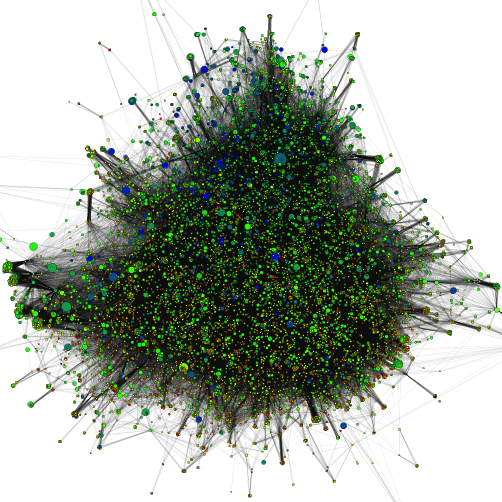}
  \caption{Complete author citation network with 8,077 nodes and 75,351 links.}\label{fig:author_citation}
\endminipage\hfill
\minipage{0.45\textwidth}
  \includegraphics[width=\linewidth]{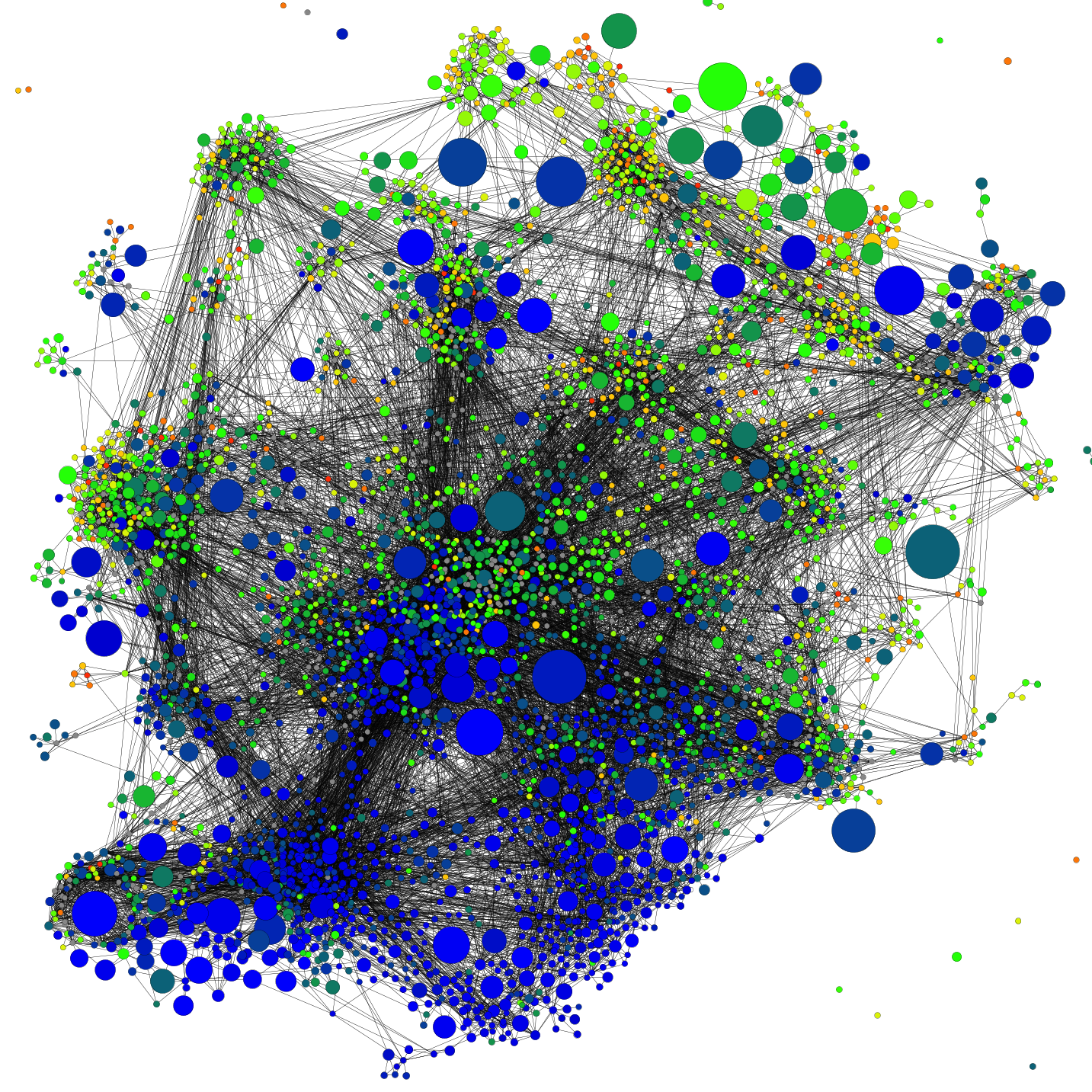}
  \caption{Complete paper citation network with 5,587 nodes and 20,660 edges.}\label{fig:paper_citation}
\endminipage\hfill \vspace*{1em}
\\

\minipage{0.45\textwidth}
  \includegraphics[width=\linewidth]{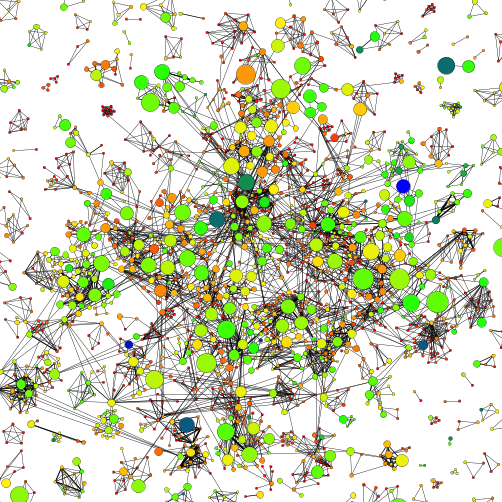}
  \caption{Complete author coauthorship network with 2,542 nodes and 5,622 edges.}\label{fig:author_coauthorship}
\endminipage\hfill
\minipage{0.45\textwidth}
  \includegraphics[width=\linewidth]{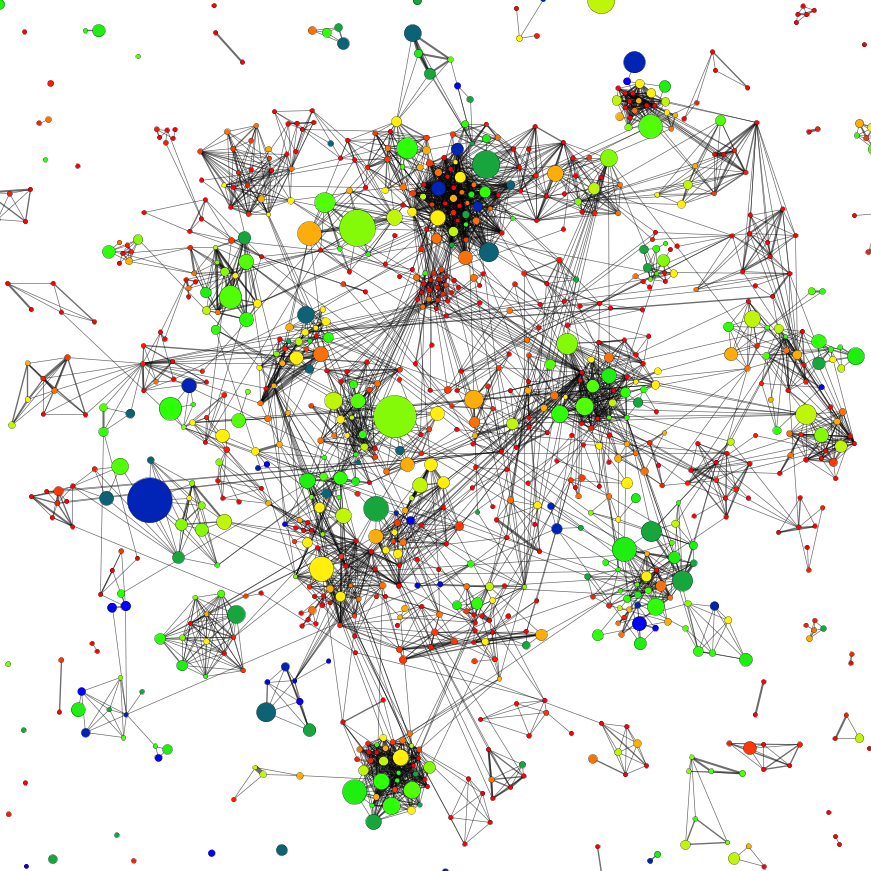}
  \caption{Complete paper coauthorship network with 1,288 nodes and 4,236 edges.  }\label{fig:paper_coauthorship}
\endminipage\hfill
\end{figure*} 

The data above encode two distinct types of relationships involving both authors and papers. Coauthorship relationships are instances of people sharing authorship of an article. Citation relationships are instances of papers referencing other papers. Because the ACM DL data contains identifiers for both types of entities, both types of relationships can be mapped between papers and authors.
\begin{description}
\item[Author-Author Coauthorship.] Nodes $i$ and $j$ are authors and an undirected link $(i,j)$ exists if they have co-authored a paper together. This link is weighted to reflect the number of papers they have co-authored.
\item[Author-Author Citation.] Nodes $i$ and $j$ are authors and a directed link $(i,j)$ exists if $i$ cites a paper written by $j$. This link can be weighted to reflect the number of times $i$ cites work by $j$.
\item[Paper-Paper Coauthorship.] Nodes $i$ and $j$ are papers and an undirected link $(i,j)$ exists if these papers share an author in common. This link can be weighted to reflect the number of authors they have in common.
\item[Paper-Paper Citation.] Nodes $i$ and $j$ are papers and a directed link $(i,j)$ exists if $i$ cites $j$. This link is unweighted because a paper can only cite another paper once. Temporal ordering necessarily makes some edges impossible: paper $i$ cannot cite paper $j$ if $i$ was published before $j$.
\end{description}

Networks were constructed and analyzed using standard methods and algorithms the NetworkX Python package~\cite{hagberg_exploring_2008}. Three types of networks were constructed. Annual networks represent the network for all the papers or authors in only a particular publication year. Complete networks represent the union of all these individual networks creating a single network across all years. CSCW sub-graphs represent only the subset of authors or papers that appeared in CSCW. Despite the presence of two distinct types of nodes (papers and authors) and methods for analyzing these bipartite networks, we convert them to one-mode networks with nodes of a single type to aid comparison and interpretation of our analyses~\cite{borgatti_network_1997}. 

Coauthorship networks are built by iterating over the papers for a given year and creating a bipartite author-paper graph linking paper IDs to author IDs. Additional meta-data such as paper or author names, years, and bibliometric data were then added as attributes of their respective nodes. Unlike the citation networks below, these coauthorship networks are only among published CSCW articles. These bipartite graphs were then projected to weighted one-mode networks of author and paper coauthorship networks. 

Because papers reference other papers outside of CSCW, citation networks were constructed using a 1.5-step ego network. The 1,288 CSCW publications formed the ``ego'' nodes and the papers that these egos referenced formed the ``alters''. For each of these alter nodes, every paper with a valid ACM DL identifier that was either referenced by the ego paper was added to an alter list and these alters' DL records were then scraped as well. These alters include papers published outside of CSCW at ACM conferences like CHI or GROUP but exclude papers in non-ACM outlets which do not receive unique IDs. Because this network construction strategy systematically excluded references to or from publications outside of the ACM DL, our analysis of citations is necessarily incomplete. The CSCW sub-graphs of the citation networks are a special class of citation networks where authors or papers are included if they have published in CSCW.

Paper citation networks were created when an ego article referenced any other ACM DL article or if an alter article referenced any other article in the set of egos and alters. Reference and citation information was not available for papers published in 2013. As a result, network structures in more recent years show dramatic changes relative to other years. To create author citation networks, directed edges were created from every author of a paper to every author of each paper cited if the alter paper was in the set of ego and alter papers. These networks capture not only whether egos are linked to alters, but also whether alters are also linked to other alters and thus retain clustering in the local network.

\section{Large-scale Structural Patterns}

\begin{figure*}[!htb]
\minipage{0.32\textwidth}
  \includegraphics[width=\linewidth]{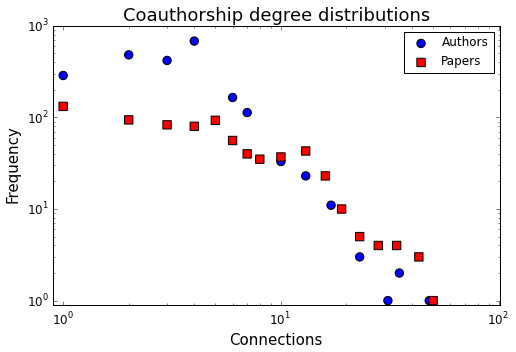}
  \caption{Degree distribution for authors (blue) and papers (red) in the complete coauthorship network.}\label{fig:coauthor_degree_dist}
\endminipage\hfill
\minipage{0.32\textwidth}
  \includegraphics[width=\linewidth]{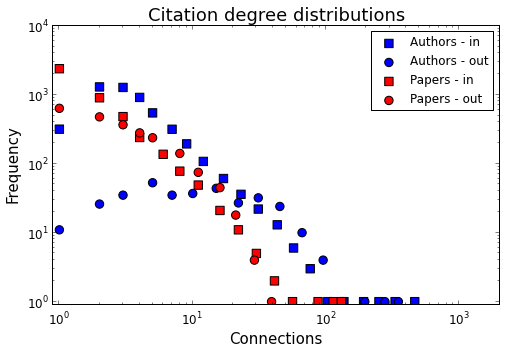}
  \caption{Directed degree distributions for authors and papers in the complete citation network.}\label{fig:citation_degree_dist}
\endminipage\hfill
\minipage{0.32\textwidth}
  \includegraphics[width=\linewidth]{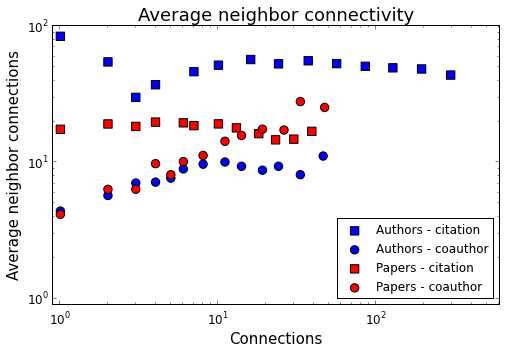}
  \caption{Average neighbor connectivity for authors (blue) and papers (red) in the citation (square) and coauthorship (circle) networks.}\label{fig:avg_connectivity}
\endminipage\hfill
\end{figure*} 

Figures~\ref{fig:author_citation} through~\ref{fig:paper_coauthorship} visualize the structure of each of the four networks. The visualizations are produced by writing the NetworkX graph objects to GraphML files  and visualizing them in Gephi graph visualization platform  using the OpenORD force-directed layout  to emphasize clusters within large-scale networks. Nodes are sized by the total number of citations received (as recorded by ACM) and are colored by year such that older nodes are bluer and younger nodes are redder. For authors, the year corresponds to the first year the author published anything indexed by ACM. For papers, the year corresponds to the year the paper was published.

Figure~\ref{fig:author_citation} is illustrative of a classic ``hairball'' network in which the network is densely tied together with indistinguishable subcomponents. CSCW sub-graphs of the author citation network are similarly dense and lack any local modularity. Authors appear to be well mixed in terms of age suggesting that newer authors are not preferentially citing other new authors. The paper citation network in Figure~\ref{fig:paper_citation} visualizes the paper-paper citation network and there is evidence of both stronger temporal homophily among papers as well as stronger sub-community structure. While old papers obviously do not have the benefit of being able to cite more recent papers, there nevertheless appears to be a lack of mixing among older papers and younger papers suggesting papers have a ``lifetime'' after which they are forgotten.

Figure~\ref{fig:author_coauthorship} visualizes the author-author co-authorship network with a focus on the largest connected component (LCC) at center surrounded by smaller, disconnected clusters. These latter clusters represent collaborations among authors on one or two papers, but these collaborations remain separate from the central CSCW community. Like the author-author citation network in Figure~\ref{fig:author_citation}, there is little evidence of strong homophily based on author age (likely an effect of advisor-advisee relationships). Similarly, older authors do not appear to have an incumbency effect that preferentially rewards them with more citations. Figure~\ref{fig:paper_coauthorship} visualizes the paper-paper co-authorship network that also has a LCC that may represent the ``core'' contributions surrounded by smaller clusters of papers sharing co-authors outside of the core. The clusters within the network represent research groups and the accumulated effort of a principal investigator pursuing a research agenda across multiple papers with several authors. Subsequent analyses will examine whether membership in the LCC of these networks confers an advantage for author and paper impact.

Figure~\ref{fig:coauthor_degree_dist} and Figure~\ref{fig:citation_degree_dist} plot the degree distributions of the complete coauthorship and citation networks respectively on log-log axes using logarithmic binning to smooth the distribution~\cite{milojevic_power_2010}. Most of the relationships show characteristic long-tailed distributions in which the majority of authors and papers have relatively few connections but some authors and papers have orders of magnitude more than others. A notable exception to this is the author out-degree citation relationship in Figure~\ref{fig:citation_degree_dist} that exhibits a flatter distribution: authors referencing one or two other authors are not significantly over-represented over authors references one or two dozen other authors.

\begin{table*}[!htb]
\centering
\begin{tabular}{cccc}
\toprule[0.125em]
Author co-authorship & Author citation & Paper co-authorship & Paper citation \\ 
\cmidrule(lr){1-4}
\centering
Robert Kraut (51) & Robert Kraut (553) & \specialcell{``The use of visual\\ information in shared\\visual spaces'' (51)} & \specialcell{``Interaction and outeraction:\\Instant messaging\\ in action'' (28)} \\
Mark Ackerman (46) & Paul Dourish (528) & \specialcell{``Action as language\\ in a shared visual\\space'' (51)} & \specialcell{``Why CSCW applications fail:\\ problems in the design\\ and evaluation of\\ organizational interfaces'' (28)} \\
Carl Gutwin (35) & Saul Greenberg (476)& \specialcell{``Social media question\\ asking workshop'' (50)} & \specialcell{``Design of a multimedia\\ vehicle for social browsing'' (21)} \\
Loren Terveen (35) & Jonathan Grudin (406) & \specialcell{``Does CSCW need\\ organization theory?'' (46)} & \specialcell{``Real time groupware\\ as a distributed system:\\ concurrency control and\\ its effect on the interface'' (21)} \\
Kori Inkpen (34) & Steve Whittaker (396) & \specialcell{``Coordination and\\ beyond: social functions\\ of groups in open\\ content production'' (40)} & \specialcell{``Operational transformation in\\ real-time group editors:\\ issues, algorithms, \\and achievements.'' (16)}\\
\bottomrule[0.125em]
\end{tabular}
\caption{Top five authors and papers by degree centrality for each complete network (number of connections).}
\label{table:top-dogs}
\vspace{-1.0em}
\end{table*}

The authors and papers with the most connections are summarized in Table 1. The authors with the highest co-authorship degree centrality have collaborated on papers with many unique authors within the CSCW community: in the case of Robert Kraut, 51 people. The authors with the highest citation degree centrality have referenced or been cited by many unique authors in the CSCW community: in the case of Robert Kraut, 553 people. The papers with the highest co-authorship degree centrality reflect the fact that its authors have written many other CSCW papers: Darren Gergle, Robert Kraut, and Susan Fussell's 2002 and 2004 papers rank highly because they have collectively authored 51 other CSCW papers. The papers with the highest citation degree centrality have referenced or been referenced by many other CSCW papers: Lynne Markus and Terry Connolly's 1990 paper and Bonnie Nardi, Steven Whittaker, and Erin Bradner's 2000 paper have both referenced and been referenced by 28 other CSCW papers.

Figure~\ref{fig:avg_connectivity} plots the average neighbor connectivity to understand patterns of assortative degree mixing across the four complete networks. For all nodes with $n$ connections (x-axis), the y-axis plots the average number of connections these nodes' neighbors have. Citation networks (squares) show flat or negative slopes (random or dissortative degree mixing) indicating that authors and papers with few references and citations connect to other authors and papers that are well connected (upper-left quadrant). Conversely, papers and authors with many references and citations are connected to authors and papers with few connections themselves (lower-right quadrant)---it is ``crowded at the bottom and lonely at the top''.  The opposite pattern holds for co-authorship networks (circles) that show positive slopes (assortative degree mixing). This indicates authors and papers with few co-authors are connected to other authors and papers with few co-authors (lower-left quadrant) while authors and papers with many connections are connected to authors and papers who well-connected themselves (upper-right quadrant)---it is ``lonely at the bottom and crowded at the top.'' Because the connectivity of one's neighbors may impact the ability to access resources, subsequent analyses examine whether average neighbor degree influences impact.

\section{Structural Changes Over Time}

\begin{figure*}[!htb]
\minipage{0.32\textwidth}
  \includegraphics[width=\linewidth]{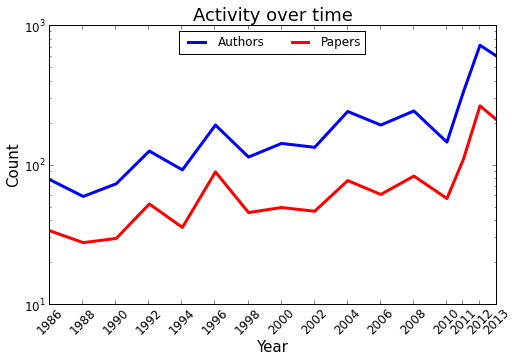}
  \caption{Number of unique authors and papers per year.}\label{fig:activity}
\endminipage\hfill
\minipage{0.32\textwidth}
  \includegraphics[width=\linewidth]{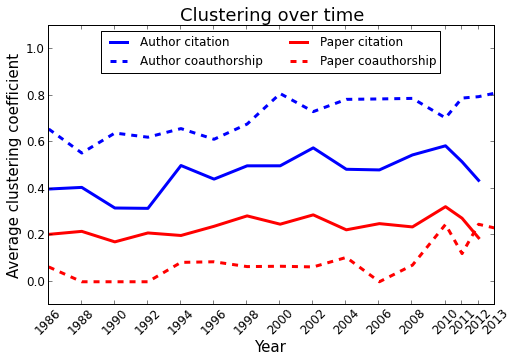}
  \caption{Clustering of the co-authorship and citation networks for authors and papers over time.}\label{fig:clustering}
\endminipage\hfill
\minipage{0.32\textwidth}
  \includegraphics[width=\linewidth]{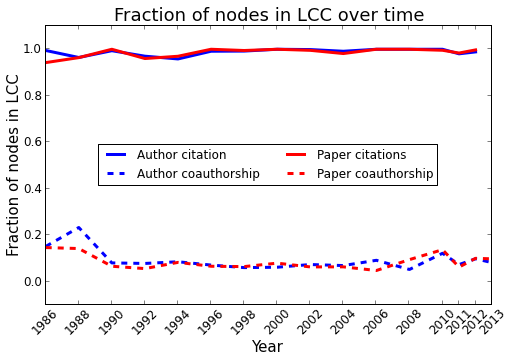}
  \caption{Fraction of nodes in the largest connected component (LCC) of the co-authorship and citation networks for authors and papers over time.}\label{fig:lcc}
\endminipage\hfill
\end{figure*}  

For an interdisciplinary conference like CSCW, balancing of cohesion and diversity of approaches is crucial. Figure~\ref{fig:activity} examines changes in activity over time. CSCW has witnessed an almost order of magnitude increase in the number of authors and papers participating in recent years compared to initial proceedings. Cohesive communities where members share connections with each other can indirectly reach many others can also lead to calcification as people become embedded and strongly constrained within methodological or theoretical sub-communities. Alternatively, sparser communities may allow people and ideas to freely move around but lacks any coherent or unifying core to attract and retain members. These concepts of cohesion and diversity can be analyzed using metrics of clustering and largest connected component activity, which we analyze for all four network types over time.

To examine the question of cohesion in more detail, we also examine the average clustering coefficients of the networks over time (Figure~\ref{fig:clustering}) and the fraction of nodes in the sub-community (Figure~\ref{fig:lcc}). Clustering generally suggests ``friends of my friends are also friends.'' Authors with high clustering have co-authors who also write papers together or reference authors who also reference each other. Papers with high clustering are connected to other papers that also share co-authors or reference papers that also reference each other. For example, some types of research may demand more resources resulting in larger collaborations centered on senior scholars who can support and manage them, creating clusters among the constituent authors and their papers. 

In Figure~\ref{fig:clustering}, authors have substantially higher clustering than papers in both relationships. Temporal censoring also affects the structure of these networks as new authors and papers have not yet had a chance to collaborate or be cited. Omitting the falloff arising from censoring in recent years' citations, there is more clustering among authors in recent years than in earlier years. Similarly, citations among papers have remained relatively stable while co-authorship has increased. This suggests authors inhabit sub-communities that tend to work closely together and reference each other's work.

The fraction of nodes in the largest connected component provides another interpretation of cohesion. The largest connected component in Figure~\ref{fig:paper_coauthorship} is at the center with some smaller isolated sub-graphs surrounding it. Networks in which all of the nodes are in the largest component means they are indirectly connected while networks in which few nodes are in the largest component suggested activity is split between many foci. In Figure~\ref{fig:lcc}, all of the cited work is in a single connected component each year and remains consistent over time. Conversely, coauthorship relations are substantially more fragmented with the main component containing consistently containing fewer than 20\% of the published authors or papers in any year. In subsequent analyses, we examine whether these measures of cohesion (clustering and membership in the giant component) influence impact.

The impact of an author or paper can be evaluated a variety of ways, but the ACM DL provides two metrics with significant ecological validity: downloads and citations. Papers that are repeatedly downloaded and cited reflect interest in reading the work and referencing its contributions in on-going scholarship. Author who have had many papers downloaded or have received many cites have analogous impacts. However, these metrics have shortcomings such as not counting downloads from non-ACM sites, censoring downloads that occurred before these metrics began to be collected in early 2002, confounding download and citation activity outside of CSCW scholarship, and excluding citations outside of ACM's indexing. Thus, while the models of impact we propose below may not generalize to total impact of an author or paper, they nevertheless capture several substantive metrics of impact within the ACM community. 

\begin{figure*}[!htb]
\minipage{0.32\textwidth}
  \includegraphics[width=\linewidth]{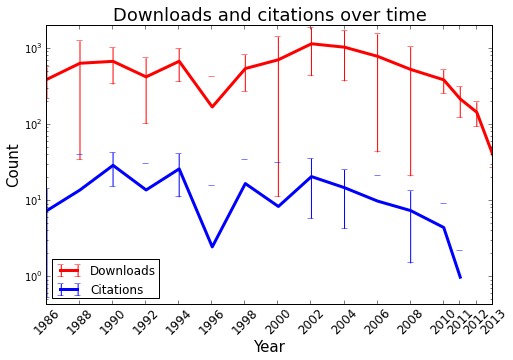}
  \caption{Changes in median (standard errors) downloads (red) and citations (blue) for papers published in each year of proceedings.}\label{fig:downloads_timeseries}
\endminipage\hfill
\minipage{0.32\textwidth}
  \includegraphics[width=\linewidth]{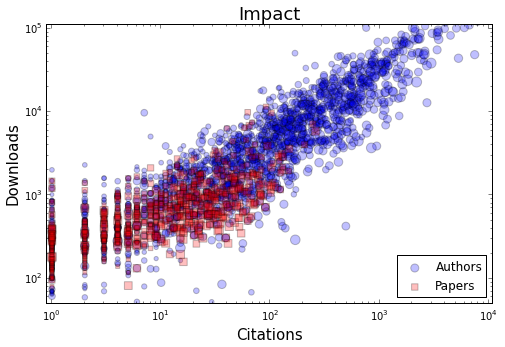}
  \caption{Scatterplot of downloads (y-axis) and citations (x-axis) metrics for authors (blue) and papers (red).}\label{fig:downloads_scatter}
\endminipage\hfill
\minipage{0.32\textwidth}
  \includegraphics[width=\linewidth]{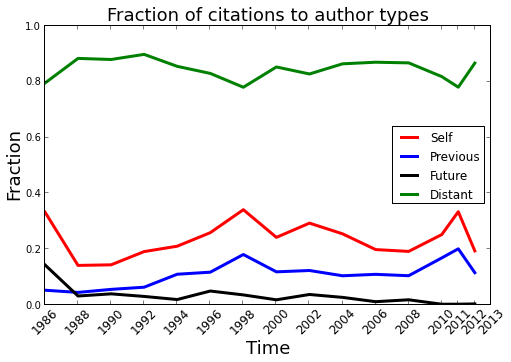}
  \caption{Fraction of papers referencing self (red), previous coauthors (blue), future coauthors (black), and distant authors (green) for papers published in each year of proceedings.}\label{fig:citation_types}
\endminipage\hfill
\end{figure*}

Figure~\ref{fig:downloads_timeseries} plots changes in the distributions of these impact factors for each year of the proceedings. There is a clear temporal censoring effect whereby publications in recent years have diminished impact owing to their novelty. Assuming more recent publications do not have systematically less impact, these distributions suggest it takes more than 5 conferences for a cohort of proceedings papers to be properly evaluated. Figure~\ref{fig:downloads_scatter} demonstrates there is a strong correlation between these impact factors across both authors and papers. 

Figure~\ref{fig:citation_types} plots changes in citation activity to authors' selves, previous co-authors, and un-related authors over time. For each of the references in each paper published in a CSCW proceedings, referenced papers that share an author with one of the CSCW paper authors are counted as self-references, referenced papers that are authored by someone whom any of the CSCW paper authors previously coauthored are previous co-authors references, and referenced papers that are authored by someone that none of the proceeding paper previously co-authored are distant references. Self-citations and citations to distant authors are relatively stable over time while there is a modest increase in the citations to previous co-authors over time. 

\begin{figure}[!htb]
\includegraphics[width=\linewidth]{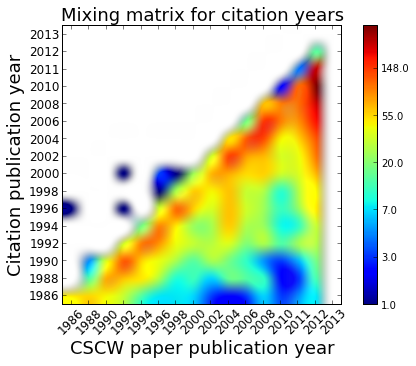}
\caption{Citation frequency for published years citing other years.}\label{fig:cross-years}
\end{figure}

Figure~\ref{fig:cross-years} is a 2D histogram of the frequencies of articles published on the x-axis year citing papers published in the y-axis year. For example, the most cited year for papers in 2012 was 2010. This figure suggests CSCW authors have a short-term memory whereby most citations to happen to papers published one to two years previously. Thereafter, the likelihood of an author citing a paper from any other year falls smoothly until approximately the 10-year mark, when citations to older papers drops off dramatically.

\section{Statistical Model of Social Capital}
We test four hypotheses relating individual paper's and author's social capital (measured by four network features) to their impact (measured by total downloads and citations). These hypotheses do not differentiate between papers or authors nor do they differentiate between coauthorship and citation relationships. As the subsequent section discusses in more detail, impact is the outcome of interest across models and is taken to be either the total number of downloads or the total number of citations.

\subsection{Hypotheses}
The first hypothesis (H1) proposes a positive relationship between brokerage and impact. For coauthorship relations, the ability to span disconnected groups gives brokers as ``vision advantage'' by having early access to less redundant and more diverse information that gives them an advantage in drawing upon a wider set of ideas and resources. In the context of citation, brokerage manifests as authors or papers referencing or synthesizing findings from different sub-communities. These brokerage positions in each types of network will lead to authors and papers to identify contributions that synthesize approaches from and make contributions to distinct sub-communities that will in turn be recognized through subsequent downloads and citations. 

The second hypothesis (H2) proposes a positive relationship between clustering and impact. Authors and papers that are strongly embedded within sub-groups of can mobilize more resources to produce research, receive expert feedback on it, and disseminate it among colleagues. In particular, highly embedded coauthorship relations often reflect co-location or mentorship that benefit from easily-exchanged resources, shared interests, and latent trust ties that facilitate the development of higher-impact research projects absent in less embedded coauthorship relations. Embedded referencing relationships benefit from shared recognition of the value of particular research topics and/or methods and these ``invisible colleges''~\cite{de_solla_price_collaboration_1970} may preferentially reward members of their group with citations and downloads.  

The third hypothesis (H3) proposes a positive relationship between neighbor connectivity and impact. Authors and papers connected to well-connected alters benefit from the association by gaining indirect access to their various forms of capital. Co-authorship with well-connected authors may grant access to more implicit and explicit resources like feedback, dissemination, recognition, and visibility that in turn lead to higher-impact projects than co-authorship with poorly-connected authors. Alternatively, citing papers that have been cited by many other scholars may signal membership in the community and valorization of the same canon of prior work that in turn leads others to recognize the author's or paper's advances over this prior scholarship.
 
The fourth and final hypothesis (H4) proposes a positive relationship between membership in the largest connected component and impact. Generally, this process should capture the tendency for authors and papers admitted to the ``core'' of the community to benefit from recognition by other elites. Membership in the core co-authorship community reflects an affiliation with elite researchers through mentorship, shared affiliation, or joint funding that provides access to more resources to develop novel projects as well as promote one's research that will lead to greater impact. Figure 10 demonstrated the vast majority of citations are in the giant component, so exclusion from this giant component by failing to cite any other paper in the CSCW canon while nevertheless being accepted to CSCW should therefore confer a major disadvantage to overall impact.

\subsection{Model specification}
To test whether impact is a function of position within the networks described above, we propose a simple linear ordinary least squares (OLS) regression model specified at the node level predicting the number of citations or downloads for papers and authors as a function of four network statistics in Model 1 through Model 8 in Table 2. These models control for variables such as number of publications published by an author and the author's tenure (first year published) as well as the number of authors on a paper and the year the paper was published. 

Citation relationships are converted from directed to undirected relationships to compute undirected statistics such as clustering and betweenness as well as to facilitate comparison between models. To correct for skewed distributions in citations, downloads, betweenness, authors, and publications, these values are log transformed. Furthermore, to facilitate the comparison of effect sizes within and between models, standardized coefficients are reported. Model selection was evaluated using the Akaike Information Criterion (AIC) with the best-fitting models minimizing the AIC.

\subsection{Results}
\begin{figure*}
\includegraphics[width=\linewidth]{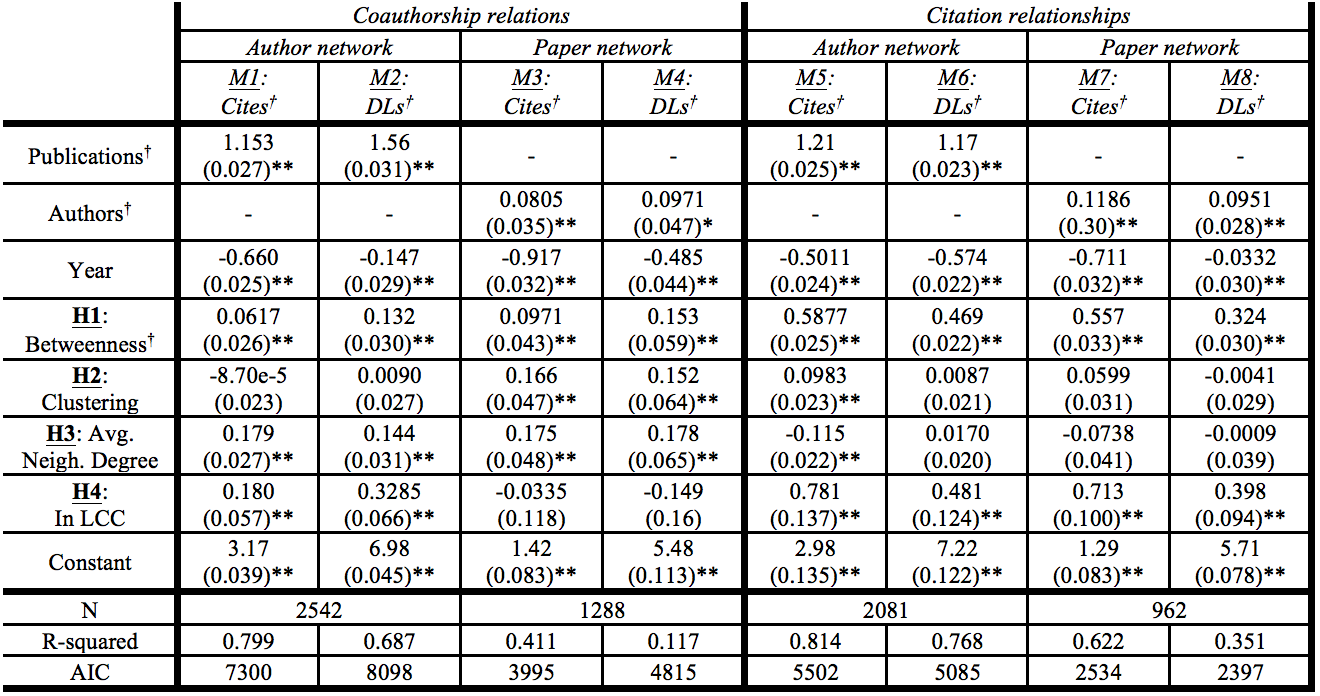}
\caption{Standardized OLS regression estimates (standard errors) for models predicting author and paper impact based on network features of coauthorship and citation networks. ? log-normalized, * p < 0.05, ** p < 0.01}\label{fig:regression}
\end{figure*}

Table 2 summarizes the standardized regression coefficients for all eight models. The model fit for author networks is particularly strong with R-squared values between 0.687 and 0.814, which is primarily driven by the publications variable. As expected, this controls for levels of activity (publications for authors and number of authors for papers) and shows strong positive effects on impact and the year of publication shows strong negative effects across all models and significant effect across all eight models. Within the coauthorship networks, the effect size of betweenness is secondary to other effects. However, gains in betweenness confer major advantages to papers and authors within their respective citation networks.. Substantively, citations connecting disparate sub-communities receive out-sized benefits to their impact suggesting the CSCW and broader ACM community reward authors and papers that synthesize approaches and thus supporting H1.

The evidence of clustering's relationship with impact is not as robust across types of relationships and nodes. The paper coauthorship network saw significant increases in paper impact among papers strongly embedded within clusters of papers sharing authors. This clustering effect for paper coauthorship implies the impact of a paper either builds upon the contributions of prior papers or the paper serves as a focal point for mobilizing authors to write additional papers to expand upon its findings. Similarly, significant and positive effects were observed in only the author citation network. Ego authors citing alter authors who cite \ other alter authors suggests the presence of a coherent community of research practice that rewards the ego authors with additional citations but not additional downloads. However, neither the author coauthorship network nor the other citation relationships show gains to impact arising from embeddedness in clusters. These results provide mixed support for H2. 

Brokerage (as measured by betweenness) shows a positive Average neighbor degree shows a strong and positive effect across coauthorship networks but minor and non-significant effects across most citation relationships, providing mixed support for H3. Citing well-connected authors appears to significantly penalize authors' citation rates but has no effects on downloads. Citing well-cited papers also has negative but non-significant effects on impact. However, this is unsurprising because citation ties are effectively free and Figure 7 already suggested a ``crowded at the bottom and lonely at the top'' phenomenon for well-cited authors and papers to cite or be cited by poorly connected authors and papers and vise versa. Substantively, this weak result suggests that impact may accrue to authors and papers that reference and thereby recognize relatively unknown papers while referencing well-known papers does not warrant recognition for oneself.

However both author and paper coauthorship relations impose greater substantive costs and the positive relationship between average neighbor connectivity and impact for these coauthorship relations is provocative. This is evidence of an ``invisible college'' of core and prolific authors who preferentially collaborate with each other. On one (optimistic) hand, the ability to recruit well-connected elites to collaborate suggests that these elites recognize the intrinsic value of an author's or paper's idea or they are able to improve co-authors' ideas better than non-elite co-authors. On the other (pessimistic) hand, the ability to author high-impact papers with elites suggests elites are able to mobilize more resources to promote their own work and may lead to a ``Matthew effect'' whereby the well-connected reinforce their position. 

Membership in the LCC is trivial for the citation networks as Figure 10 demonstrated nearly every node is a member of the LCC. As such, the strong effects observed in M5 through M8 reflect the penalties assigned to authors and papers that fail to reference any authors or papers in the giant component. However coauthorship networks, where the majority of activity occurred outside the LCC, had more complex results providing mixed support for H4. Membership in the LCC for author co-authorship networks had a very strong and positive effect for both types of impact while membership in the LCC for paper co-authorship networks had negative and non-significant estimates. Substantively, this suggests there are major impact benefits that accrue to members of the community who are able to co-author with other core authors. In light of the findings from H3, this is potentially problematic as it suggests some individuals' impact is further limited by their ability mobilize core members of the community to co-author with them. However, the core papers in CSCW confers no similar advantage suggesting paper impact emerges from other processes like cumulative development captured by clustering in H2. 

To review, occupying brokering positions in all four types of relationships showed statistically significant and major effects on impact, supporting H1. Other network features of social capital show more mixed results. Average neighbor degree showed a consistently positive results for coauthorship networks suggesting impact is strongly influenced by the ability to mobilize the resources of elite coauthors but had negligible effects for citation patterns, partially supporting H3. LCC membership had strong effects in the citation network suggesting that recognition within the community demands recognizing others in the community but little effect for coauthorship networks, partially supporting H4. While clustering conferred strong and significant advantages to impact for paper coauthorship, there was little evidence of embeddedness affecting impact in other relationships, providing weak evidence for H2. 

\section{Discussion}
Our findings show that CSCW's coauthorship and citation networks have highly-skewed levels of participation that are found in many other bibliographic analyses of scholarly communities. Analysis of changes in these structures over time points to the stability of these patterns despite significant shifts in the thematic focus, methodological orientation, and other structural features like the frequency and size of the conference as well as the length of contributions and changes in the submission process~\cite{grudin_computer-supported_1994,grudin_taxonomy_2012,olson_groupware_2003}. Taken together, these empirical findings are indicative of a cohesive and stable scholarly community that is able to generate findings that are valued outside of the community, socialize newcomers into productive members over several generations, and remain resilient to major changes in technologies, social relationships, and the organization of the conference itself. However, the lack of substantive shifts in the size, distribution, or clustering of collaborations among members results are also somewhat surprising in light of the falling costs of obtaining data and remote communication and increasing availability of systems for managing distributed work. 

We also tested a statistical model of scholarly impact based upon four network features of social capital. The findings from these models suggest that the impact of an author or paper as measured by downloads and citations can be predicted very well based entirely on the patterns of co-author and referencing relationships rather than features intrinsic to the paper itself. In general, brokering between otherwise disconnected groups is a strong predictor for both authors and papers across co-authorship and citation relationships. Co-authoring papers with prolific authors appears to confer a substantial advantage to impact for both papers and authors. Papers earn additional impact if their authors have co-authored many other papers but not from being members of the core community. The opposite holds for authors' impact as they benefit from being members of the core community but need not co-author within highly clustered groups.

Taken together, these network features of social capital as they relate to scientific impact have very problematic implications for the CSCW community. Our findings are consequential because they imply the value of good ideas emerge primarily from the structural position within the network, rather than the idea itself or the person articulating it. Authors at the periphery of these networks or who lack co-authors in the core appear to be systematically ignored by the broader community while authors in the core or with well-connected co-authors have their advantage preferentially reinforced. The devaluation of peripheral authors and their contributions potentially does serious injury to the idea that CSCW is an interdisciplinary research venue if recognition within the broader ACM community is contingent upon membership or affiliation with elite members of the community. In other words, although the number of ``handshakes'' one is from Robert Kraut, Jonathan Grudin, or Judy Olson should not determine whether the CSCW community values your research, it nevertheless does appear to matter. 

Of course, these elite scholars are not personally responsible for this outcome: the download and citation data are two quantitative metrics that are drawn from aggregated activity of thousands users across the entire ACM community. Members of a community like CSCW should be able to demarcate the boundaries of a field's research agenda, theories, and methods so to exclude pseudo-scientific approaches and ensure the quality necessary to maintain the legitimacy of the field for the purposes of recruitment, promotion, and funding~\cite{gieryn_boundary-work_1983}. Although recent changes to the submission and revision system clearly have the potential to address the biases our analysis has identified, the implications of these changes are too recent and the data are too sparse to analyze with any confidence. Nevertheless, our findings suggest that even aftering overcoming the substantial barriers to entry from the submission and review process, papers and their authors face an uphill battle to attract the attention of a community that appears content to reward attention on elites.

In the interests of ``eating our own dogfood'' we used the coauthorship models in Table 2 to project the impact of this paper by 2024 by inserting it into the author and paper co-authorship network models (M1-M4). The paper network model based upon co-authorship predicts this paper will receive 637 downloads and 9.98 citations by 2024. 

\subsection{Future work}
The results from our analysis are largely descriptive but open the door for a variety of follow-on studies to understand processes of CSCW collaboration through the lens of CSCW. First, while collaborations may bring about cross-fertilization of ideas from distinct domains, the costs and risks of collaboration may also lead to retrenchment around established focal research projects. The success of brokerage in these models suggests that CSCW remains a small world rife with local clustering whose structural holes are waiting to be spanned. we identified brokerage as an important mechanism for both authors and papers to generate impact, it is unclear which, if any, of Burt's forms of brokerage we captured: making the brokered parties aware of shared interests, transferring best practices, identifying analogies between previously unrelated information, and finally synthesizing new information from disconnected groups~\cite{burt_structural_2004}.

Second, the particular social processes that govern the accumulation of impact are unclear. Diffusion processes through word-of-mouth recommendation or pedagogical replication, activation processes such as timeliness of new or old findings, inheritance processes of others citing the author or paper and the work being discovered thereafter are all possibilities. Our paper primarily examined structural features, but textual features and labels could begin to unpack the substance authors articulate through these documents and how they diffuse throughout the community. Modeling of keyword labels for papers could help understand the changing research interests of a field and its members. In particular, the ability to bring together authors with diverse keyword backgrounds may contribute to impact under a variety of logics such as brokering position, tackling larger problems, or moving quickly to be the first to write on a subject. Keywords could also be used to trace authors' trajectory through a field and trace their evolution as growing in specialization or diversifying across disciplines. Speculative research that is either extremely novel (as measured by keyword diversity) or extremely timely (as measured by being among the first to mention a keyword) likely predicts future impact.

Third, despite the longitudinal data available, the statistical analysis we used was cross-sectional across all years. While this approach reflects the fact that scholarship is both archived and cumulative, it nevertheless limits our ability to make causal claims about the processes that contribute to scholarly impact. Additional research could be adapted to model the future research productivity of authors given their prior work, the influence of credit on a paper (e.g., name order), and the persistence or lifetime of collaborations between authors. The opens the potential of scholarly ``sabermetrics'': identifying talented researchers whose work is being undervalued or overlooked. Furthermore, publications are not the only outputs of scholarly collaboration nor are citation and co-authorship the only relations within scientific collaborations. Some strong collaboration or influences are not ``consummated'' into a paper or citation while some weak collaborations and ideas do manifest as co-authorships and citations. Networks based on institutional or disciplinary affiliations could reveal other networked structures about authors' backgrounds and proximity, which also likely mediate patterns of collaboration. Authors can also be bound together by mentorship, acknowledgements, and attendance at internships, symposia, or workshops.

Fourth, a variety of distinctive processes such as homophily, closure, balance, and resource dependence can explain the creation of a tie. Statistical network methods like p*/exponential random graph models (p*/ERGMs) could be used to estimate multilevel statistical models of how endogenous network processes of closure and centrality, exogenous node attributes such as gender and age, and exogenous dyadic covariates such as citation and coauthorship relations influence the structuring one these networks~\cite{contractor_testing_2006}. Other statistical network models to capture temporal dynamics of the graph evolving in more detail to understand co-evolutionary processes of how changing co-authorship and citation links and changing node attributes of impact, age, or gender influence one another~\cite{snijders_introduction_2010}.

Fifth, given our findings that social position within the networks of CSCW explains over 80\% of the variance in scores, the community should pursue additional research and discussions about how to address systematic biases in the submission and review process. In addition there is also potentially substantial institutional and geographic biases in submission rejection rates and the patterns of gender, institutional, and geographic representation on the program committees that may contribute to these biases.

Finally, there is a dearth of reflexive scholarship about the extent to which CSCW scholars use CSCW technologies to support their own research collaborations. Interviews and surveys of authors could be used to understand their motivations in selecting collaborators and identifying research topics about which to write. Recommender systems might even be developed to maximize the potential impact of a paper by mining these data to identify where the greatest opportunities for scholars to collaborate to produce high-impact work. Furthermore, the practice of collaboration and how researchers use tools like e-mail, instant messaging, online document editors, code repositories, and social media to support co-authorship collaborations as well as to document scholarship for citation and disseminate information to their social circles is also a rich domain for understanding the interactions between co-authorship, citation, social capital, and impact. 

\section{Conclusion}
The age, size, interdisciplinarity, and influence of the CSCW community make it an attractive subject for analyzing the structure and dynamics of its collaborations. Drawing upon methods in bibliometrics and network science we analyzed how these structures have changed since its foundation and the implications that positions within these structures have for broader research impact. Testing a variety of network theories of social capital, we estimated a statistical model relating network position within co-authorship and citation networks to impact metrics like citations and downloads. This model suggested the impact of CSCW authors and papers are strongly correlated with their position within these collaboration and citation networks and that brokerage positions for both authors and papers consistently confer substantial advantages to impact. Other network features of social capital have more complex relationships with impact across networks but, network position nevertheless explains a remarkable amount of variance in author and paper impact. These findings have problematic implications for the community as it suggests impact does not stem from intrinsic values of papers or authors but from their relationships. We conclude by outlining a research agenda for future work to examine the social processes that confer this advantage.

\section{Acknowledgements}

%
%
%
%
%
\balance

\bibliographystyle{acm-sigchi}
\bibliography{bibliometrics}
\end{document}